\documentclass[aps,prb,preprint, superscriptaddress, draft, showpacs]{revtex4}
\usepackage[final]{graphicx}

\usepackage{hyperref}
\graphicspath{{./figs/}}
\begin{document}

\title{Correlation between microstructure and magnetotransport in organic semiconductor spin valve structures}

\author{Yaohua Liu}
\affiliation{Dept. of Physics and Astronomy, The Johns Hopkins University, Baltimore, MD, 21218}
\author{Shannon M. Watson}
\affiliation{NIST Center for Neutron Research, National Institute of Standards and Technology, Gaithersburg, Maryland 20899, USA}
\author{Taegweon Lee}
\affiliation{Dept. of Materials Science and Engineering, The Johns Hopkins University, Baltimore, MD, 21218}
\author{Justin M. Gorham}
\affiliation{Dept. of Chemistry, The Johns Hopkins University, Baltimore, MD, 21218}
\author{Howard E. Katz}
\affiliation{Dept. of Materials Science and Engineering, The Johns Hopkins University, Baltimore, MD, 21218}
\author{Julie A. Borchers}
\affiliation{NIST Center for Neutron Research, National Institute of Standards and Technology, Gaithersburg, Maryland 20899, USA}
\author{Howard D. Fairbrother}
\affiliation{Dept. of Chemistry, The Johns Hopkins University, Baltimore, MD, 21218}
\author{Daniel H. Reich}
\email[Corresponding author, ]{reich@jhu.edu}
\affiliation{Dept. of Physics and Astronomy, The Johns Hopkins University, Baltimore, MD, 21218}

\date{\today}

\begin{abstract}
We have studied magnetotransport in organic-inorganic hybrid multilayer junctions. In these devices, the organic semiconductor (OSC) Alq$_3$ (tris(8-hydroxyquinoline) aluminum) formed a spacer layer between ferromagnetic (FM) Co and Fe layers. The thickness of the Alq$_3$ layer was in the range of 50-150~nm. Positive magnetoresistance (MR) was observed at 4.2~K in a current perpendicular to plane geometry, and these effects persisted up to room temperature. The devices' microstructure was studied by X-ray reflectometry, Auger electron spectroscopy and polarized neutron reflectometry (PNR). The films show well-defined layers with modest average chemical roughness (3-5~nm) at the interface between the Alq$_3$ and the surrounding FM layers. Reflectometry shows that larger MR effects are associated with smaller FM/Alq$_3$ interface width (both chemical and magnetic) and a magnetically dead layer at the Alq$_3$/Fe interface. The PNR data also show that the Co layer, which was deposited on top of the Alq$_3$, adopts a multi-domain magnetic structure at low field and a perfect anti-parallel state is not obtained. The origins of the observed MR are discussed and attributed to spin coherent transport. A lower bound for the spin diffusion length in Alq$_3$ was estimated as $43 \pm 5$~nm at 80~K. However, the subtle correlations between microstructure and magnetotransport indicate the importance of interfacial effects in these systems.

\end{abstract}

\pacs{72.25.Dc, 72.80.Le, 85.75.-d}

\maketitle

\section{Introduction}

There is currently a rapidly increasing interest in spin-dependent electronic transport in organic semiconductors. At its heart is the expectation that weak spin-orbit coupling in these light-element-based materials will lead to long spin relaxation times and long spin coherence lengths that may ultimately enable their use in magnetoelectronic devices. In addition, the wide range of organic semiconductors (OSCs) and the ability to tune their properties by suitable chemical modifications holds promise for increased flexibility in controlling spin injection and matching interface properties in multilayered devices such as spin valves. Although it has been generally accepted that there is a conductivity mismatch problem for spin injection from ferromagnetic (FM) contacts into semiconductors,~\cite{Schmidt2000} this problem can be potentially solved via a spin-dependent interfacial barrier,\cite{RashbaPRB2000, SmithPRB2001, RudenSmithJAP2004} which might be either a Schottky barrier or an insulating interface layer.~\cite{SmithPRB2002, MinNatmat2006, SantosPRL2007, DediuPRB2008} Experimentally, large magnetoresistance (MR) effects were originally reported in FM1/OSC/FM2 trilayers, where the bottom ferromagnetic layer was La$_{0.67}$Sr$_{0.33}$MnO$_{3}$ (LSMO), the OSC was tris(8-hydroxyquinoline) aluminum (Alq$_3$), and the top FM layer was Co.\cite{XiongNature2004} Such effects have since been observed for a range of OSCs~\cite{WangPRB2007, OsterbackaAPL2006} and in devices where both FM layers are transition metals.\cite{WangSynthMetal2005, PramanikPRB2006, SantosPRL2007}

However, the nature of the spin transport in these multilayer devices remains unclear, and both spin coherent transport~\cite{XiongNature2004, WangPRB2007, OsterbackaAPL2006, WangSynthMetal2005} and spin polarized tunneling~\cite{SantosPRL2007, XuAPL2007} have been invoked to explain the observed MR. Furthermore, it is challenging to distinguish among competing theories as reproducibility of sample quality has been an issue in many experimental studies in organic spintronic devices.~\cite{PettaPRL2004, PramanikPRB2006, IkegamiAPL2008} Samples grown by different groups do not consistently show large MR effects,~\cite{WangSynthMetal2005, JiangPRB2008} presumably due to subtle structural variances.  Particularly problematic has been the question of the degree of interdiffusion of the top FM layer into the much softer OSC layer during sample fabrication.~\cite{XiongNature2004, Vinzelberg2008} Thus experiments that relate directly the physical and magnetic structure of FM1/OSC/FM2 systems with the existence or non-existence of large MR effects in such structures are essential. In particular, X-ray reflectometry (XRR)~\cite{FullertonPRL1992, SchadPRB1998} and polarized neutron reflectometry (PNR),~\cite{Fitz2004} with their ability to probe the structural and magnetic properties of buried interfaces, are well suited to this task. Here we report XRR and PNR results, in conjunction with SQUID magnetometery, Auger depth profiling, and electrical transport studies to correlate the microstructure and magnetotransport in OSC spin valve structures.

Our patterned devices used Fe and Co as the top and bottom FM electrodes and Alq$_3$ with thickness from 50~nm to 150~nm as the spacer layer. Positive magnetoresistance (MR) was observed at 4.2~K and these effects persisted up to room temperature. The microstructure was studied by XRR, Auger electron spectroscopy (AES), and PNR on a series of unpatterned films that were co-deposited with the samples used for the transport and magnetization studies. These films showed well-defined layers with modest average chemical roughness (3-5~nm) at the interface between the Alq$_3$ and the surrounding FM layers. By comparing samples with similar Alq$_3$ thicknesses, but with different magnetotransport properties, we found direct correlation between the MR amplitude and the microstructure at the FM/Alq$_3$ interfaces. The magnetic interface is generally smoother than the chemical interface at the Fe/Alq$_3$ boundary. Larger MR effects are associated with smaller FM/Alq$_3$ interface widths and with a magnetically dead layer at the Alq$_3$/Fe interface. Such magnetically dead interfaces may circumvent the resistance mismatch problem~\cite{RudenSmithJAP2004} and facilitate effective spin injection.

\section{Sample Fabrication}

Co/Alq$_3$/Fe samples were prepared in thermal evaporation chambers with a base pressure of 2~$\mu$Torr. The films were deposited on Si wafers with 300~nm SiO$_2$ top layers. Commercially purchased Alq$_3$ was purified by sublimation before evaporation. The bottom FM film (typically 25 nm Fe) was first deposited at ambient temperature. For transport samples, a shadow mask was used to define the Fe layer into rectangles $6 \times 0. 8$~mm$^{2}$. Without breaking vacuum, the shadow mask was removed to deposit Alq$_3$. The Alq$_3$ thickness was varied from 50~nm to 150~nm.  The vacuum chamber was then opened to change shadow masks to permit fabrication of cross junctions for transport measurements. A second FM layer consisting of 5~nm Co and a 40~nm Al capping layer was then deposited in a rectangle $6 \times 0. 8$~mm$^{2}$ with its long axis perpendicular to the Fe strip.  The Fe and Co were deposited at a rate of 0.2~nm/s, the Al was deposited at 0.4~nm/s and the Alq$_3$ was deposited at 0.1~nm/s. In order to stabilize the OSC films during evaporation and to reduce the potential penetration of Co atoms into the OSC, chilled water was used to keep the substrate holder at 20$^{\circ}$C during Co and Al deposition. The Co layer was kept thin to limit further the increase in sample temperature during the Co deposition.~\cite{XiongNature2004,WangPRB2007,WangSynthMetal2005,XuAPL2007} The active junction area for the transport samples was $800 \times 800$~$\mu$m$^{2}$ and the unpatterned samples used for reflectivity studies had dimensions $16 \times 16$~mm$^{2}$. The film thickness was monitored by a quartz crystal oscillator during evaporation and the actual thickness of each layer was determined from the reflectivity experiments.

\section{Experimental Methods}

A standard four probe configuration was used for the magnetotransport measurements. The measurements were carried out in a continuous flow cryostat in the temperature range from 4~K to 290~K. An in-plane magnetic field in range of $\pm 200$~mT was applied parallel to the top Co electrode to control the relative magnetization direction of the top and bottom FM layers. Magnetic hysteresis loops of the films were obtained using a SQUID magnetometer at a series of temperatures. The films' structure was studied by X-ray reflectometry, using a 4-circle diffractometer. The X-ray source uses a ceramic filament tube with a Cu-target. $\Omega$-$2\theta$ scans were performed, where $\Omega$ is the angle between the incident beam and the sample surface, and $2\theta$ is the scattering angle. A pair of Soller slits with divergence 0.04$^{\circ}$ were used in the beam path to limit the axial divergence of the X-ray beam. A curved graphite monochromator was used to select the Cu $K_{\alpha}$ radiation. In a typical specular reflectivity scan, $2\theta$ varied from 0$^{\circ}$ to 5$^{\circ}$, with resolution better than 0.01$^{\circ}$. Background data were obtained using a $0.1^{\circ}$ offset in $\Omega$. Elemental depth profiling was carried out using a Scanning Auger Microprobe. Auger electrons were collected in three energy windows for a series of sputtering cycles. In each cycle, the sample was sputtered for 2 minutes with 1 keV Ar ions and then electrons were collected as a function of kinetic energy with energy step 1~eV over the ranges 745~eV to 790~eV, 245~eV to 295~eV, and 560~eV to 610~eV. These ranges isolate the Co LVV, C KLL, and Fe LMM characteristic peaks, respectively.

To identify the devices' spin structure, we employed PNR techniques~\cite{MajkrzakPhysicaB1991, Fitz2005} to determine the depth-dependent vector magnetization of individual FM layers on sub-nanometer length scales, using the NG-1 polarized neutron reflectometer at the NIST Center for Neutron Research. For the PNR experiments, neutrons were polarized parallel to the applied field in the sample plane. A vertically-focusing pyrolytic graphite monochromator was used to select neutrons with wavelength of 4.75~\AA ~and a wavelength divergence of 0.05~\AA. The neutron beam had an angular divergence $\delta \theta = 0.018^{\circ}$. By employing two spin flippers, all four spin scattering cross sections were measured, including two non-spin flip (NSF) reflectivities, $R^{(++)}$ and $R^{(--)}$, and two spin flip (SF) reflectivities, $R^{(+-)}$ and $R^{(-+)}$. Within the context of kinematic theory, the four cross sections for a perfectly polarized beam are described by:~\cite{MajkrzakJAP1988, MajkrzakPhysicaB1991, Fitz2005}

\begin{eqnarray}\label{eq:PNR}
R^{(\pm\pm)}(Q)&\propto &|\int [ \rho(z) \pm C M(z) \sin\phi(z)] e^{iQ_{Z}Z}dz|^2, \\
R^{(\pm\mp)}(Q)&\propto &|\int C M(z) \cos\phi(z) e^{iQ_{Z}Z}dz|^2,
\end{eqnarray}

where $\rho$ is the nuclear scattering length density (related to the chemical composition), $M$ is the in-plane magnetization, and $\phi$ is the angle between $M$ and the applied field $H$. $C = 2.9109 \times 10^{-5}$~\AA$^{-2}$T$^{-1}$ is a constant connecting the magnetic moment density with the magnetic scattering length density. For the NSF reflectivities, the neutron retains its original polarization after scattering from the sample, and for the SF reflectivities the neutron changes spin states. The NSF reflectivities provide information concerning the chemical composition of the film and are sensitive to the component of the in-plane magnetization aligned along the field axis. The SF reflectivities are sensitive only to that component of the in-plane magnetization perpendicular to the field direction.

Specular reflectivity scans were made over the range of $0^{\circ} \leq 2\theta \leq 5.2^{\circ}$. For non-specular reflectivity scans (used for background subtraction), $\Omega = 0.5 \times 2\theta - 0.3^{\circ}$ and $2\theta$ covered the same range as used in the specular scans. Rocking curves were taken by rotating the sample plane from $\Omega = -1.4^{\circ} $ to $2.8^{\circ}$ while fixing the detector at $2\theta = 1.4^{\circ}$. The PNR measurements were taken at both 40~K and room temperature. At each temperature, PNR was measured at two fields, a high field at which the average in-plane magnetization of the Co and Fe were parallel and a low field at which they were antiparallel. These fields were chosen based upon characteristics of magnetic hysteresis loops measured with SQUID magnetometery. Both the X-ray and neutron specular reflectivity data were corrected for instrumental background, efficiencies of the polarizing elements (neutrons only, typically $> 98\%$), and the footprint of the beam. The Reflpak software suite, which uses a least squares optimization, was used for elements of the XRR and PNR data reduction and analysis.\cite{reflpak}

\section{Results and Analysis}
\label{sec:RA}
\subsection{Magnetotransport measurements}

\begin{figure}[th]
	\centering
		\includegraphics[width=0.48\textwidth]{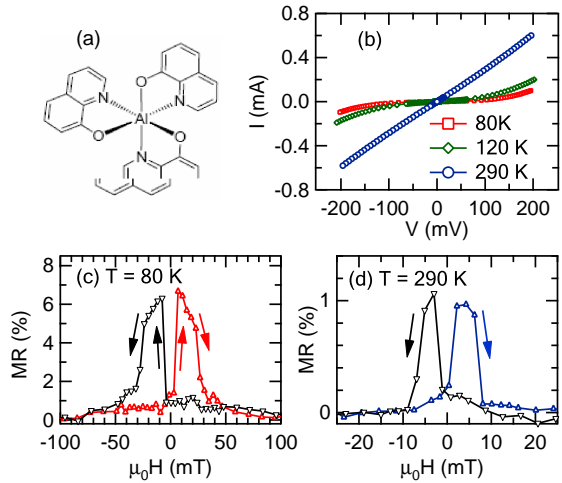}
	\caption{\label{MR}(a) Chemical structure of Alq$_3$. (b) $I$-$V$ curves of a junction with 64~nm Alq$_3$ spacer layer, taken at 80~K, 120~K and 290~K. The $I$-$V$ curves show increased nonlinearity and increasing junction resistance as the temperature decreases. (c) and (d) show the magnetoresistance data of this device taken at 80~K and 290~K, respectively. The data were taken with a four probe configuration, and a DC current $I = 1$~$\mu$A. The high field resistances were 22.2~k$\Omega$ at 80~K and 355~$\Omega$ at 290~K.}
\end{figure}

The chemical structure of Alq$_3$ is shown in Fig.~\ref{MR}a. Figure~\ref{MR}b shows current-voltage ($I$-$V$) curves for a junction with a 64~nm thick Alq$_3$ layer. These are distinctly nonlinear at low temperatures. As the temperature increases, the $I$-$V$ curves become more linear, and at the same time both the junction resistance and the MR decrease. Alq$_3$ is a well known electron carrying OSC.~\cite{XiongNature2004, ZhanPRB2008} The junction was biased so that electrons were injected from Fe into Alq$_3$ at positive bias and from Co into Alq$_3$ at negative bias. The $I$-$V$ curves show very weak asymmetry for positive bias and negative bias. The maximum MR observed for this junction was about 9\% at 80~K and 1\% at 290~K, as shown in Fig.~\ref{MR}c and Fig.~\ref{MR}d, respectively. Both the nonlinear $I$-$V$ curve and the temperature dependence of the junction resistance exclude the possibility of a metallic short between the two FM layers in this sample.~\cite{WangPRB2007, AkermanAPL2000}

\begin{figure}[th]
	\centering
		\includegraphics[width=0.45\textwidth]{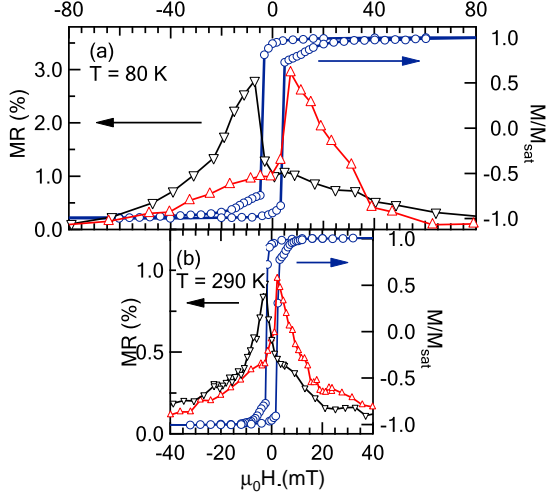}
	\caption{\label{MR_MH} (a) Magnetoresistance data (triangles) and the hysteresis loop (blue circles) for a sample with Alq$_3$ thickness of 85~nm. Both data sets were taken at 80~K with the magnetic field parallel to the Co strip. The magnetoresistance shows a jump at the same field where the magnetization along the field changes its sign, i.e. at $\sim \pm 5$~mT. (b) Similar data taken at 290~K. The sudden changes of the magnetoresistance and magnetization occur at lower fields, $\sim \pm 2.5$~mT, than at 80~K. The measurement current was 10 $\mu$A and the high field resistance was 640~$\Omega$ at 80~K and 25~$\Omega$ at 290~K.}
\end{figure}

To illustrate the relationship between the magnetic switching of the Co and Fe layers and the observed MR of these structures, a magnetic hysteresis loop is shown overlaid with magnetotransport data in Fig.~\ref{MR_MH}a for a junction with an 85~nm Alq$_3$ layer. Both data sets were taken at 80~K with the magnetic field parallel to the Co strip. This sample showed a maximum 3\% MR at 80~K. The hysteresis loop shows a two-step switching of the layers' magnetization. When the field was swept from $-200$~mT to 200~mT, the net magnetization abruptly changed sign at $\sim 5$~mT. This large change arises from the magnetization reversal of the Fe layer, which is approximately three times as thick as the Co layer. As the field is increased above 5~mT, the magnetization gradually approaches saturation. Above 80~mT, the magnetization is essentially saturated. The sample shows a rise in MR that correlates with the nominal Fe magnetization reversal at $\sim 5$~mT, followed by decreasing MR with increasing field. When $\mu_0 H > 80$~mT, no apparent hysteresis in either the magnetization or the MR is observed. Figure~\ref{MR_MH}b shows similar data taken at 290~K. Here, the magnetization reversal happened at a lower field, $\sim \pm 2.5$~mT. This is characteristic behavior of a spin valve device where the high and low resistance states come from the configurations with antiparallel and parallel alignment of the magnetization of the FM layers, respectively. The shift to lower field is presumably due to a decrease in the barrier to switching with increasing temperature.

\begin{figure}[th]
\includegraphics[width=0.48\textwidth]{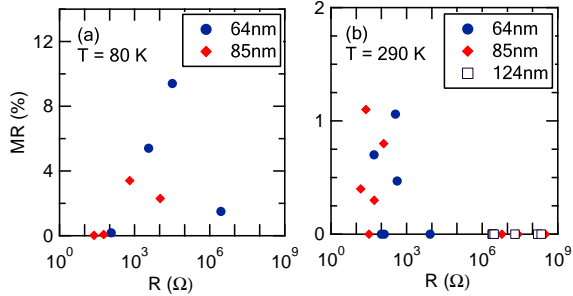}
\caption{\label{RvsMR} Maximum MR measured in 3 sets of junctions with different Alq$_3$ thickness as a function of the junction resistance, data taken at (a) 80~K and (b)  290~K, respectively. The maximum MR observed in a set of junctions decreases as the Alq$_3$ thickness increases. However, there is large variability in both junction resistances and MRs from the same batch.}
\end{figure}

Magnetotransport properties were measured for a series of junctions with Alq$_3$ thicknesses from 50~nm to 150~nm. For each set of junctions with the same Alq$_3$ thickness, magnetotransport measurements were performed on 6-12 samples. The results are summarized as maximum MR observed in a sample as a function of the sample's resistance for 3 sets of junctions, as shown in Fig.~\ref{RvsMR}. These junctions are labeled by the thickness of the Alq$_3$ layer. Both room temperature and 80~K results are shown. The set of junctions with 124~nm Alq$_3$ generally have resistances larger than 1~G$\Omega$ at 80~K, which limits our ability to make accurate transport measurements. We found significant variability in both resistance and MR, consistent with what has been observed in other FM/OSC/FM devices up to now.~\cite{PettaPRL2004, PramanikPRB2006, IkegamiAPL2008} Although we confirmed the tendency that the MR increased with decreasing thickness of the Alq$_3$ layers, the large variability makes it difficult to obtain the effective spin diffusion length accurately from the thickness dependence of the MR.

Despite this, the spin diffusion length in Alq$_3$ can still be estimated using the Julliere model,~\cite{Julliere1975} under the assumption that there is no spin scattering at the FM/OSC interfaces and the injection current has the same spin polarization as the bulk FM leads due to the so-called self adjustable interface effect.~\cite{XiongNature2004, PramanikNatNano2007} In this calculation, we further neglect effect of interface intermixing and take the thickness of the OSC layer as the actual spin transport length.~\cite{PramanikNatNano2007, ShimPRL2008} The sum of the interface roughnesses at both FM/OSC interfaces was taken as the uncertainty of the transport length in the Alq$_3$ layer, which is about 8 nm from our XRR study (see Sec.~\ref{sec:XRR}). The bulk spin polarizations for Co and Fe are 45\% and 43\% respectively.~\cite{StrijkersPRB2001} Therefore, from the maximum MR of 9\% observed for the junction with 64~nm Alq$_3$, we obtain a spin diffusion length of $\lambda_S = 43 \pm 5$~nm at 80~K, which is similar to the value reported by Xiong \emph{et al.}~\cite{XiongNature2004} However any spin scattering at the interface or formation of magnetic multidomain structure will decrease the MR value and will result in an underestimation of $\lambda_S$. Hence our estimation sets a lower bound for the spin diffusion length in Alq$_3$.

\subsection{X-ray reflectivity}
\label{sec:XRR}
\begin{figure}[th]
\includegraphics[width=0.4\textwidth]{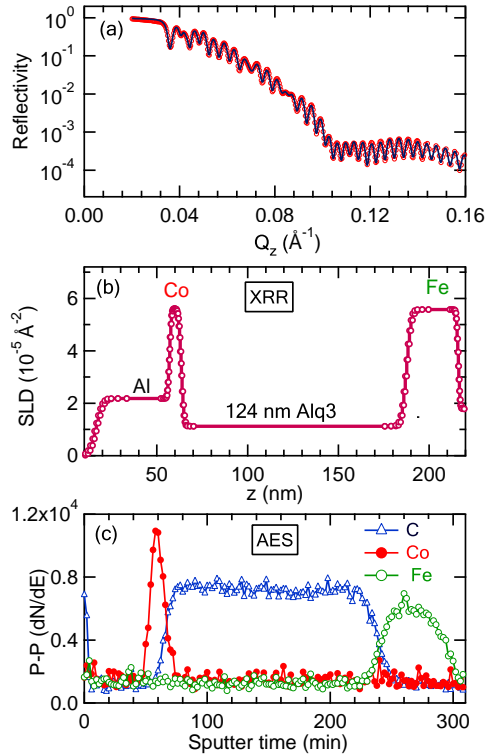}
\caption{\label{XRRAES}(a) Specular XRR data (dots) together with the fit (solid line) for a Al/Co/Alq$_3$/Fe multilayer. (b) The SLD profile acquired from the fit, which shows well defined layer structures. This result is confirmed by the elemental depth profile acquired by AES (c) on a sample from the same batch. In these data, the carbon signal indicates the presence of Alq$_3$.}
\end{figure}

X-ray reflectivity was used to study the structure of the spin valve films.  An example of data obtained in such measurements is shown in Fig.~\ref{XRRAES}a.  The data are shown vs the momentum transfer perpendicular to the film plane $Q_z = 4\pi \sin{\theta}/\lambda$. The data shown in Fig.~\ref{XRRAES}a are dominated by oscillations with period $\Delta Q \approx 0.004$~\AA$^{-1}$. This corresponds to the total thickness of the Co/Alq$_3$/Fe trilayer, which is about 160~nm. However, additional modulations of the scattering are also present, which enable full determination of the structural depth profile.

The reflectivity data were fit using the Reflpak software system,~\cite{reflpak} which employs Parratt's algorithm~\cite{MajkrzakSPIE1992}
to model interface roughness as a sequence of very thin slices whose scattering length density (SLD) and absorption vary smoothly so as to interpolate between adjacent layers in the spin-valve stack.  The SLD profile determined in this manner for this sample is shown in Fig.~\ref{XRRAES}b, and the best-fit curve for the reflectivity data is overlaid on the data in Fig.~\ref{XRRAES}a.  The principal results determined from these fits are that there are well-defined regions of the film corresponding to the Co, Alq$_3$, and Fe layers, respectively, and that the interfacial roughnesses are small, typically 3-5~nm, indicating that there is only limited mixing between the OSC and FM layers.  These results also yield accurate measurements of the individual layer thicknesses.  In this case, they are $t_{Co} = 6.44 \pm 0.05 $~nm, $t_{Alq_3} = 124.05 \pm 0.05 $~nm, and $t_{Fe} = 28.51 \pm 0.05$~nm.

To confirm the accuracy of the XRR results, the elemental composition of a second film co-deposited with that shown in Figs.~\ref{XRRAES}a and~\ref{XRRAES}b was measured by AES depth profiling. In AES, the peak to peak height of the count rate differentiated with respect to the energy $dN/d \epsilon$ is proportional to the concentration for each probed element. These results are shown in Fig.~\ref{XRRAES}c, where the differential peak to peak heights are plotted as a function of sputtering time. The average sputtering rate was approximately 0.7~nm/min.  The vertical resolution of AES is limited principally by the inelastic mean free path (IMFP) of the ejected electrons. For all of the elements (C, Co and Fe) measured in Fig.~\ref{XRRAES}c, the IMFPs are on the order of $1-2$~nm. The effect of sputtering is weak since there is only a slight asymmetry as the depth profile passes through each elemental region. Therefore, the resolution in our experiment is roughly same as the step size, which was on average 1.5~nm. No attempt has been made to convert sputtering time into depth due to the considerable differences in sputtering rates of the materials being probed, notably two metals (Fe and Co) and an organic (Alq3), which makes a quantitative conversion from sputter time to depth ambiguous.~\cite{ElecSpec1994} Thus, in essence the AES shows that the Fe/Alq3/Co film consists of discrete multilayers, whose thicknesses are in qualitative agreement with the predictions of the SLD fit.

\begin{figure}[th]
\includegraphics[width=0.4\textwidth]{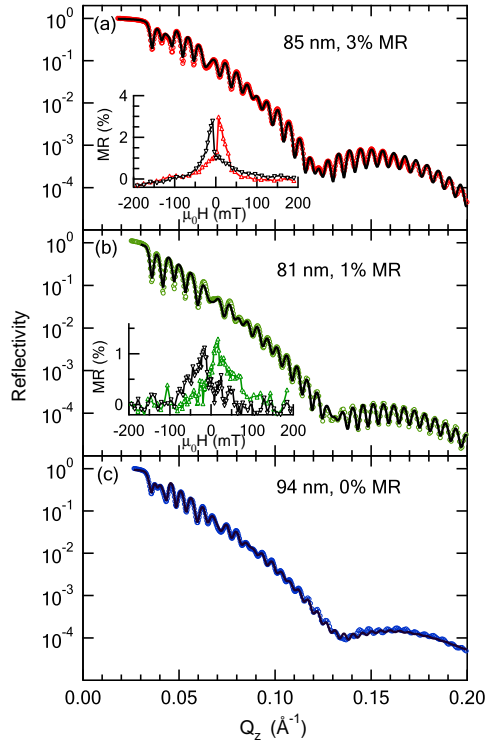}
\caption{\label{XRR_MR} XRR data and fits for three films with comparable Alq$_3$ thicknesses, (a) 85~nm , (b) 81~nm and (c) 94~nm, respectively. The insets in (a) and (b) show the magnetotransport data on junctions codeposited with these films taken at 80~K (a) and 10~K (b). The maximum MR observed was 3\%, 1\% and 0\%, respectively. }
\end{figure}

\begin{figure}[th]
\includegraphics[width=0.48\textwidth]{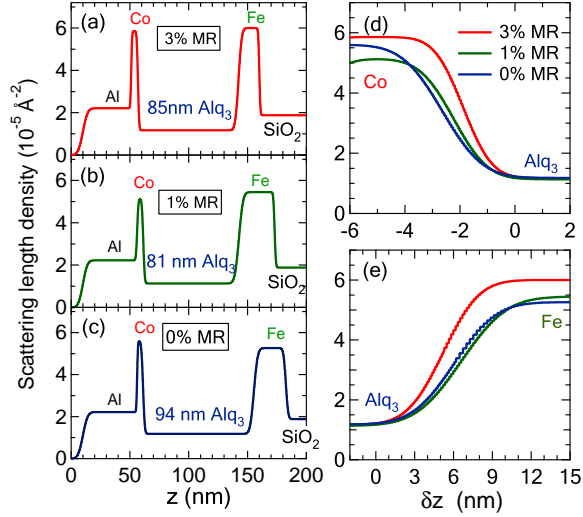}
\caption{\label{XRR_RO} (a)-(c) SLD from fitting the XRR data for three samples with similar Alq$_3$ thicknesses but different MRs. (d) and (e) Details of the structural roughnesses at the interfaces between the FM metals and Alq$_3$. Offsets in z were applied to align the curves at the FM/Alq$_3$ interfaces. $\delta z$ is defined as the distance with respect to a point near the Alq$_3$ side, where $QC_{nor}(z)= 0.01$. $QC_{nor}(z) = (QC(z)- QC_{Alq_3})/(QC_{FM}- QC_{Alq_3})$, where $QC_{Alq_3}$ and $QC_{FM}$ are the fit values of the SLDs for the Alq$_3$ layer and the appropriate FM layer.}
\end{figure}

Figure~\ref{XRR_MR} shows XRR data for three films with comparable Alq$_3$ thicknesses: 85~nm, 81~nm and 94~nm, respectively. Samples with similar Alq$_3$ thicknesses were selected in order to control for possible effects associated with Alq$_3$ thickness, which allowed us to focus on the interfacial effects on the magnetotransport properties. The insets in Fig.~\ref{XRR_MR} show magnetotransport data on junctions co-deposited with these films, which show maximum MR of 3\% (Fig.~\ref{XRR_MR}a), 1\% (Fig.~\ref{XRR_MR}b) and 0\% (Fig.~\ref{XRR_MR}c). Hereafter, we will label these samples as the 3\%, 1\% and 0\% MR samples. (The 3\% MR sample is the same one as that shown in Fig.~\ref{MR_MH}.) The depth dependence of the SLD for the three samples as determined from the fits to the reflectivities is plotted in Fig.~\ref{XRR_RO}a-c. The best-fit Co SLDs are about 10\% lower than the bulk values and varied from sample to sample in a range of 10\%. For the 3\% MR sample, the best-fit Fe SLD is bulk-like, but for the 1\% and 0\% MR samples, it is $\sim$ 7\% lower than the bulk value, possibly indicative of voids or strain which reduce the average density. The measured SLD of the Alq$_3$ layers $QC_{Alq_3} = 1.16 \times 10^{-5}$~\AA$^{-2}$ and varies by less than 2\% from sample to sample. Details of the SLD depth profiles at the interfaces between the FM layers and the OSC are shown in Fig.~\ref{XRR_RO}(d)-(e). In order to compare the interfacial variation of SLDs among different samples, we introduced the parameter $\delta z$, which is the depth z with an offset (see the caption for definition) in order to align the data at the FM/Alq$_3$ interfaces. A clear correlation between the maximum observed MR and the widths of the FM/Alq$_3$ interfaces was observed,  with sharper interface width corresponding to larger MR. The fits are more sensitive to the structure at the Co/Alq$_3$ interface than to that at the Alq$_3$/Fe interface. The widths (10\% - 90\%) are 2.1~nm, 2.5~nm, and 2.9~nm at the Co/Alq$_3$ interface and 5.7~nm, 7.2~nm, and 6.4~nm at the Alq$_3$/Fe interface for the 3\% MR, 1\% MR and 0\% MR samples, respectively.

\subsection{Polarized neutron reflectivity}
Studies on metallic spin valve structures have shown that the chemical structure alone is not sufficient to determine magnetotransport properties.~\cite{Fitz2004} We therefore employed PNR to explore the depth-dependent magnetic structure in these samples. PNR was measured for each sample at 40~K and 290~K, and at two fields for each temperature. The measurement fields were chosen based on magnetic hysteresis loops taken at each temperature, such as those shown in Fig.~\ref{MR_MH} at 80~K and 290~K for the 3\% MR sample. The high field PNR data for all samples were taken at 200~mT, where the magnetization of both the Co and Fe layers were saturated parallel to $H$. The low field state was prepared by sweeping the field from -200~mT to a small positive field just above the abrupt increase in $M$($H$) where the Fe magnetization appears to reverse direction (e.g., Fig.~\ref{MR_MH}). For the 3\% MR sample at 40~K, the Fe layer switching occurred at $\mu_0H \simeq 8$~mT, and the PNR measurements were made at 10~mT.

\begin{figure}[th]
\includegraphics[width=0.45\textwidth]{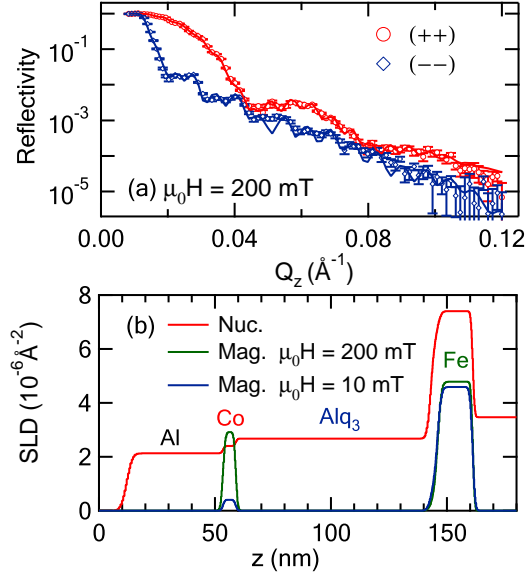}
\caption{\label{PNR_S925}(a) High field (200~mT) specular reflectivity data (circles) with fits (solid lines) of the two NSF cross sections, $R^{(++)}$ and $R^{(--)}$, for the 3\% MR sample. The data were taken at 40~K. (b) The nuclear and magnetic SLDs obtained from the fits.}
\end{figure}

\begin{figure}[th]
\includegraphics[width=0.48\textwidth]{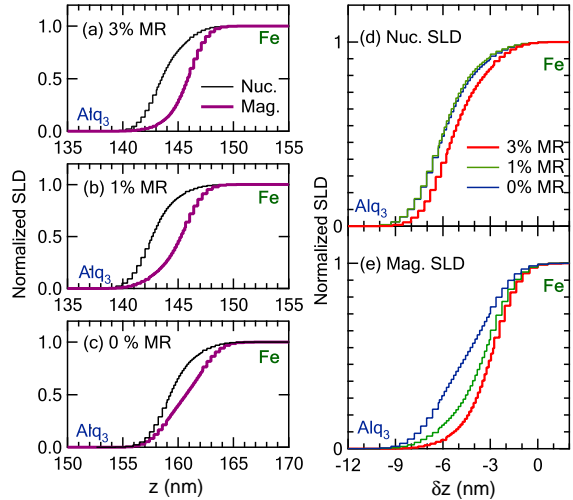}
\caption{\label{PNR_Fe}Normalized nuclear and magnetic SLDs at the Alq$_3$/Fe interface for (a) 3\% MR, (b) 1\% MR and (c) 0\% MR samples, respectively. These values are obtained for data taken at 40~K and 200~mT. Different depth dependences of the nuclear and magnetic SLDs were observed at the Alq$_3$/Fe interface. The normalized nuclear SLD $QN_{nor}$(d) and the normalized magnetic SLD $QM_{nor}$ (e) for all three samples. $\delta z$ is defined as the distance with respect to a point near the Fe side, where $QM_{nor}(z)= 0.99$. The sample showing the largest MR has the sharpest magnetic interface, while the difference in the nuclear SLDs is much smaller.}
\end{figure}

The PNR data obtained in a high saturating field for all three samples are typified by the data shown for the 3\% sample in Fig.~\ref{PNR_S925}a. All of our specular SF reflectivity scans had negligible scattering (data not shown), which indicates no net moment aligned perpendicular to the applied field. There is a clear splitting between the $R^{(++)}$ and $R^{(--)}$ cross sections, which indicates that, in addition to the nuclear scattering, there is significant scattering from the in-plane magnetization, as can be seen from Eq.~\ref{eq:PNR}. In order to obtain a reasonable fit to the high-frequency oscillations in the $R^{(--)}$ cross section, it was necessary to divide the Fe layer into two sections with independently-varying values of the magnetic SLD. The  nuclear and magnetic SLD profiles determined from the fits are shown in Fig.~\ref{PNR_S925}b. The magnetic SLD does not track the nuclear SLD as the magnetized region of the Fe layer does not fully extend to the nuclear Alq$_3$/Fe interface.  In order to see the difference more clearly, we plot the normalized nuclear and magnetic SLDs for each of the three samples in Fig.~\ref{PNR_Fe}a-c, using a definition similar to that used for the XRR data.  For direct comparison, the normalized nuclear SLDs at the Alq$_3$/Fe interface for all the three samples are plotted in Fig.~\ref{PNR_Fe}d along with the normalized magnetic SLDs in Fig.~\ref{PNR_Fe}e. The trends in the nuclear SLDs are similar to those identified from the X-ray reflectivity data (Fig.~\ref{XRR_RO}) with the 3\% sample showing the sharpest Alq$_3$/Fe interface. Since the magnetic hysteresis data (Fig.~\ref{MR_MH}) indicate that the magnetization is saturated in this field of 200~mT, we conclude that the difference between the nuclear and magnetic SLD for each sample (Fig.~\ref{PNR_Fe}a-c) corresponds to a region of the Fe layer near the Alq$_3$/Fe interface in which the magnetization is suppressed.  The observation of a magnetically dead layer is not a total surprise and has been reported for several magnetic multilayer structures.~\cite{CablePRB1986, PechanJAP1994, HoffmannPRB2005} It is notable that the sample showing the largest MR has the largest magnetic dead region near the Alq$_3$/Fe interface (Fig.~\ref{PNR_Fe}e).

\begin{figure}[th]
\includegraphics[width=0.45\textwidth]{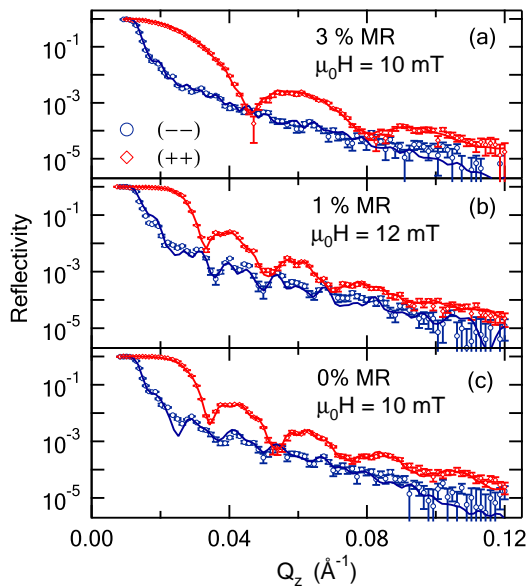}
\caption{\label{LFPNR} Low field NSF PNR data and best fits for the (a) 3\%, (b) 1\% and (c) 0\% MR sample, respectively. Data were taken at 40~K.}
\end{figure}

The low-field reflectivity data at 40~K (Fig.~\ref{LFPNR}) show features, especially in the $R^{(--)}$ cross section, that differ substantially from those in the high-field data (e.g., Fig.~\ref{PNR_S925}a). The differences between the high field and low field data originate from the change of the magnetic state. Fits to the data for all three samples (Fig.~\ref{LFPNR}a-c) reveal that the magnetically dead layer near the Alq$_3$/Fe interface persists (e.g., Fig.~\ref{PNR_S925}b), and that the Co layers' moments are aligned antiparallel to the Fe layers' moments, as expected from the magnetization curves.  However, while  the fitted moment of the Co layers for the 0\% and 1\% MR samples nearly matches the Co saturation moment (Table ~\ref{Co_SLD}), the damped oscillations in the $R^{(--)}$ data for the 3\% MR sample (Fig.~\ref{LFPNR}a) are best fit with a model in which the Co magnetization, averaged across the sample plane, is greatly reduced from its bulk value. This small, net Co moment is aligned antiparallel to the Fe moment (Table ~\ref{Co_SLD}).

\begin{figure}[th]
\includegraphics[width=0.4\textwidth]{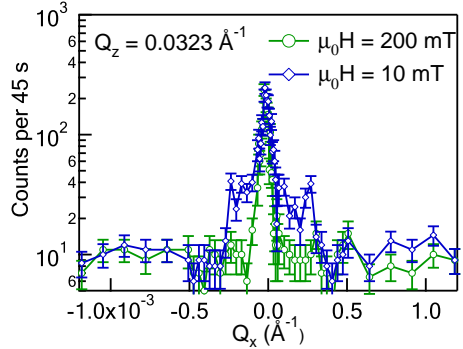}
\caption{\label{SF_S925}Spin flip rocking curves of the 3\% MR sample taken at 200~mT and 10~mT, respectively. The data were taken at 40~K and $Q_z$ = 0.0323~\AA$^{-1}$. The enhanced scattering at low field  is consistent with the observation of the much smaller Co magnetization along the field direction at low field (Table~\ref{Co_SLD}). These observations strongly suggest magnetic multi-domain formation at low field.}
\end{figure}

\begin{table}[ht]
	\centering
		\begin{tabular}
		  {|l|c|c|c|}
          \hline
                     & 3\% MR   & 1\% MR & 0\% MR \\ \hline
          High Field &  2.9     & 3.4    & 2.8   \\ \hline
          Low Field  &  -0.4    & -2.0   & -2.8   \\ \hline
		\end{tabular}
	\caption{Summary of high field and low field in-plane magnetic SLDs of the Co layer at 40~K for 3 different samples. The data are in units of 10$^{-6}$~\AA$^{-2}$. The minus sign of the SLDs at low field indicates that the magnetization of the Co layer is antiparallel to the external field.}
	\label{Co_SLD}
\end{table}

The reduction of the Co magnetization for the 3\% sample is accompanied by magnetic diffuse scattering in the SF rocking curves, as shown in Fig.~\ref{SF_S925}. The diffuse scattering disappears upon application of a saturating magnetic field of 200~mT, indicating that it likely originates from in-plane magnetic domains.~\cite{BorchersPRB1996, BorchersPRL1999} While some diffuse scattering was observed in low fields for the 1\% and 0\% MR samples, this scattering is most pronounced for the 3\% sample. Contrary to the expectation for an ideal spin valve, perfect antiparallel alignment of the Co and Fe moments is thus not achieved in the sample that exhibits the maximum MR.  Instead, our PNR results suggest that the Co layer breaks up into multiple magnetic domains within the sample plane at low field.~\cite{BorchersPRB1996, BorchersPRL1999}

\begin{figure}[th]
\includegraphics[width=0.45\textwidth]{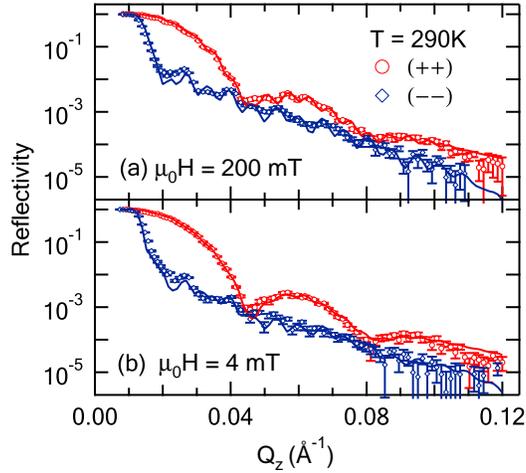}
\caption{\label{S925_290K}Specular reflectivity data (circles) of two NSF cross sections taken at 290~K for 3\% MR sample with fits (solid lines). (a) High field 200~mT and (b) low field 4~mT.}
\end{figure}

The 290~K PNR data in high and low fields appear superficially similar to the 40~K data. Figure~\ref{S925_290K} shows typical data for the 3\% MR sample. However, in comparison to the 40~K data (Fig.~\ref{PNR_S925}a), the splitting between the $R^{(++)}$ and $R^{(--)}$ specular reflectivities is slightly smaller at 290~K. The best-fit values for the magnetic SLD (Table~\ref{S925_PNRtemp}) show that this difference originates mostly from a decrease of the in-plane magnetization of the Fe layer. The reduction of the saturation magnetization of the Fe layer from 40~K to 290~K is actually much larger than that expected for a thick, bulk Fe film. The magnetic SLD of the Co layer at low field is very small for all three samples and, in general, the fits at low field are much less sensitive to the orientation of the Co magnetic moment relative to the Fe layer. The best fits indicate that the Co layers are aligned parallel to the Fe layers in all three samples at low fields.  These PNR results are consistent with the observed reduction of the MR at 290~K relative to low temperature.

\begin{table}[ht]
	\centering
		\begin{tabular}
		  {|c|c|c||c|c|}
          \hline
                      & \multicolumn{2}{c||}{Co}              & \multicolumn{2}{c|}{Fe}                  \\
          \cline{2-5} & \small~High field~ &\small~Low field~ &\small~High field~ &\small~Low field~     \\ \hline
   \small 40~K    &  2.9               & -0.4             & 4.8               & 4.6                  \\ \hline
   \small 290~K   &  2.9               & 0.8              & 4.6               & 4.4                  \\ \hline
		\end{tabular}
	\caption{The in-plane magnetic SLDs of the Co and the Fe layers for the 3\% MR sample at different experimental conditions. The data are in units of $10^{-6}$~\AA$^{-2}$. }
	\label{S925_PNRtemp}
\end{table}

\section{Discussion}

The thickness of the Alq$_3$ layer in our samples ranges from 50~nm to 150~nm, which is far above the range where tunneling would be expected to contribute to the magnetotransport. The resistance of the junctions increases quickly as temperature decreases, which argues against the existence of metallic pinholes.\cite{AkermanAPL2000} Our XRR and AES results suggest that there is no large-scale intermixing between the FMs and the Alq$_3$. It is possible that cobalt atoms that penetrate into the Alq$_3$ could behave as dopants due to charge transfer from Co to Alq$_3$.\cite{ZhanPRB2008} However it is unlikely that the MR behavior is caused by the tunneling of chains of cobalt nanoclusters for our samples, as suggested by Vinzelberg \emph{et al.}~\cite{Vinzelberg2008} The XRR data show that the concentration of cobalt decays by a factor of 9 over a distance of 2.1~nm at the Co/Alq$_3$ interface for the sample showing large MR. Therefore assuming an exponential decay of the cobalt concentration, there is approximately 1 cobalt atom for 10$^{18}$ atoms inside the Alq$_3$ matrix at a distance of 40~nm below the cobalt layer. The chance of  finding a cobalt cluster of a few~nm size is thus very small, and formation of a conductive path of cobalt chains over distance more than 60~nm is highly unlikely. The XRR results show that large MR is also associated with small chemical roughness at the Co/Alq$_3$ interface. This observation also suggests that the MR is not caused by cobalt chains, because greater roughness would result in a higher probability of forming such chains.

The observed MR is also not caused by anisotropic magnetoresistance (AMR) of the FM leads. We used a four probe configuration, and hence the contribution of the lead resistance is minimized. Furthermore, the resistance of the junction is much larger than that of the leads. The measured amplitude of the AMR of the Fe lead is much smaller than the MR of the junction. Changing the external field direction from parallel to the Co strip to parallel to the Fe strip does not significantly change the shape of the MR curve for the junction, but it causes the AMR curves of the Fe leads to change sign. Because the Co layer is in parallel with a much thicker nonmagnetic Al layer, four probe magnetotransport measurements alone did not detect any appreciable AMR ($\ll 0.1 \%$).

Possible contributors to the temperature dependence of the $I$-$V$ curves include both the injection barriers or the bulk resistance of the Alq$_3$. A satisfactory analytic theory for charge injection at a metal-organic interface and subsequent transport in an organic semiconductor is still lacking.~\cite{ScottVacSci2003} However, if we assume the effective resistance of Alq$_3$ layer is much smaller than the effective resistance caused by the injection barrier, which is a desired condition for spin injection from FM metals into semiconductor,~\cite{RashbaPRB2000, SmithPRB2001, RudenSmithJAP2004} then the injection barrier can be estimated either by the Brinkman-Dynes-Rowel (BDR) model~\cite{BDRJAP1970} for an asymmetric injection barrier or by the Simmons model~\cite{SimmonsJAP1963, SimmonsJAP1964} for a symmetric barrier. Both the BDR and Simmons models yield a temperature dependent barrier height ($\Phi$) and width ($s$), but we find that BDR model fits the data slightly better.

From the BDR model, we obtained $\Phi = 0.21$~eV, $s = 5.1$~nm, and an asymmetry $\delta\Phi = -0.17$~eV at 80 K. These values change to $\Phi = 0.27$~eV, $s = 4.3$~nm, and $\delta\Phi = -0.16$~eV at 120~K. The barrier height is lower and the barrier width is larger when compared to devices based on LSMO leads.~\cite{XuAPL2007, WangPRB2007} However they are similar to the values found on other FM/OSC structures employing transition metal FMs. For example, Santos \emph{et al.}~\cite{SantosPRL2007} and Shim \emph{et al.}~\cite{ShimPRL2008} found that a smaller barrier height and larger barrier width, due to insertion of an ultrathin Al$_2$O$_3$ layer, result in significantly enhanced spin injection from transition metal FMs into OSCs. The origin of this low and wide injection barrier in our junctions is not clear yet, however it is likely related to the detailed microstructure at the FM/OSC interfaces. We also consider the effect of the resistance of the organic layer. This semiconducting layer can be viewed as a thick barrier with a small barrier height. It is in series with the resistive injection barrier. It is essential that only a small fraction of the bias voltage drop occurs across the semiconducting layer for the observation of large MR in the FM/SC hybrid system.~\cite{RashbaPRB2000} However fitting the $I$-$V$ curve, without taking account of the OSC resistance will generally result in an overestimation of the barrier thickness and an underestimation of the barrier height.

The $I$-$V$ curves show weak asymmetries for positive and negative biases, as shown in Fig.~\ref{MR}b. From the fits, we found that the barrier asymmetries $|\delta\Phi| < 0.2$~eV. Negative values of the asymmetries suggest that electrons are more easily injected into Alq$_3$ from the Fe layer than from the Co layer. Based on the work function of pristine metals fabricated and measured in ultrahigh vacuum conditions, it was expected that there would be a larger difference in injection barriers,  $\sim$ 0.5~eV, between Fe and Co.~\cite{JiangPRB2008} However, the work functions of both Fe and Co in contact with OSCs decrease and the difference is only about 0.1~eV when fabricated in a vacuum similar to what we used here.~\cite{CambellAPL2007} This is close to our finding, and may explaine the observation of the weak asymmetry.

As temperature increases, the fits show that the injection barrier height increases and the barrier width decreases. At the same time, the MR decreases. In order to observe a large MR, both FM/OSC interfaces have to be effective for spin injection.~\cite{RashbaPRB2000, RudnerPRL2007} This requires that both interfaces are spin selective and have large effective resistances compared to the spacer layer. Therefore the temperature dependence of the effective resistances of  both injection barrier and spacer layer will certainly affect temperature dependence of the MR.  Unfortunately, the simple tunneling models can not identify the these two contributions separately. More detailed analysis is still needed, including studies of the bias dependencies of the effective resistances, however it is beyond the scope of the present study.

Our XRR/PNR study shows that the observation of large MR is directly related to the sharpness of the chemical interface between the FM layers and the Alq$_3$ layer, and also to the sharpness of the magnetic interface between the Fe and the Alq$_3$. In metallic spin valve systems, the same tendency has been observed and attributed to enhanced spin flip scattering due to rough interfaces.~\cite{SchadPRB1998} Additional interesting results come from the observation of the sharper magnetic interface as compared to the chemical interface at the Alq$_3$/Fe boundary. Larger MR is correlated to the presence of a magnetically dead region near the Alq3/Fe interface which acts as a barrier between the bulk Alq$_3$ layer and the magnetic Fe layer which acts a polarized electron reservoir. This magnetically dead interface could behave as a spin injection barrier which helps to circumvent the resistance mismatch issue at the FM/OSC interface.~\cite{RudenSmithJAP2004, SantosPRL2007, DediuPRB2008} At the same time, we observed that as the temperature increases, the in-plane magnetization of Fe in saturation field decreases much more than expected. This may contribute partially to the reduction of observed MR as temperature increases.

A reduced magnetic moment at the interface in FM/NM multilayer structures has been found in several different systems before.~\cite{CablePRB1986, PechanJAP1994, HoffmannPRB2005} While our current experiment has not pinned down the origin of the magnetic dead layer, both roughness and impurities could contribute to the reduced magnetization. Further study of the interfacial magnetization of these high Curie temperature transition metal films adjacent to OSCs would be worthwhile.

The fits to the PNR are less sensitive to the Co magnetic SLD because this layer is much thinner than the Fe layer. However we do find that there is a large reduction of the magnetic SLD of the Co layer at low field as well as strong magnetic diffuse scattering for the sample showing the largest MR. These observations suggest that thin Co films on top of Alq$_3$ are likely to form multi-domain magnetic structures at low fields and a poorly defined anti-parallel state. This most likely explains the gradual change in MR with increasing field and the lack of a well-defined plateau in the MR in the high resistance state. We estimated the spin diffusion length to be $\lambda_S = 43 \pm 5$~nm from the maximum MR of 9\% observed for a junction with 64~nm Alq$_3$. Under the same assumption, we obtain $\lambda_S = 33 \pm 3$~nm for the 3\% sample used in the structural studies. The discrepancy is likely from the decreased antiparallel alignment at low field for the set of 3\% samples, which reduces the MR and results in an underestimation of $\lambda_S$.

\section{Conclusion}

Our work demonstrates that it is possible to use transition metal FMs to inject spin polarized current into OSCs at room temperature. However, samples fabricated with similar conditions do not show consistent magnetotransport properties. We find that this originates from subtle differences in the microstructure of the samples, especially at the FM/Alq$_3$ interfaces. Larger MR is  associated with smaller structural roughness at the Co/Alq$_3$ and Alq$_3$/Fe interfaces, and  with a magnetically dead region near the Fe/Alq$_3$ interface.  The fact that significant MR can be observed in such samples even when an ideal antiparallel state is not formed implies  an important role of the interfaces in controlling spin injection from FM metals to OSCs. In addition, the magnetically dead layer we have observed may  circumvent the resistivity mismatch problem. Further studies of interfacial effects, especially on the interfacial magnetization and the injection barrier, are needed for these organic systems.

\begin{acknowledgments}
This work was supported by NSF Grant No. DMR-0520491. We acknowledge the support of the National Institute of Standards and Technology, U.S. Department of Commerce, in providing the neutron research facilities used in this work. The AES measurements were carried out in the surface analysis laboratory at Johns Hopkins University.

\end{acknowledgments}

\bibliography{OSV}

\begin{thebibliography}{44}
\expandafter\ifx\csname natexlab\endcsname\relax\def\natexlab#1{#1}\fi
\expandafter\ifx\csname bibnamefont\endcsname\relax
  \def\bibnamefont#1{#1}\fi
\expandafter\ifx\csname bibfnamefont\endcsname\relax
  \def\bibfnamefont#1{#1}\fi
\expandafter\ifx\csname citenamefont\endcsname\relax
  \def\citenamefont#1{#1}\fi
\expandafter\ifx\csname url\endcsname\relax
  \def\url#1{\texttt{#1}}\fi
\expandafter\ifx\csname urlprefix\endcsname\relax\def\urlprefix{URL }\fi
\providecommand{\bibinfo}[2]{#2}
\providecommand{\eprint}[2][]{\url{#2}}

\bibitem[{\citenamefont{Schmidt et~al.}(2000)\citenamefont{Schmidt, Ferrand,
  Molenkamp, Filip, and van Wees}}]{Schmidt2000}
\bibinfo{author}{\bibfnamefont{G.}~\bibnamefont{Schmidt}},
  \bibinfo{author}{\bibfnamefont{D.}~\bibnamefont{Ferrand}},
  \bibinfo{author}{\bibfnamefont{L.~W.} \bibnamefont{Molenkamp}},
  \bibinfo{author}{\bibfnamefont{A.~T.} \bibnamefont{Filip}}, \bibnamefont{and}
  \bibinfo{author}{\bibfnamefont{B.~J.} \bibnamefont{van Wees}},
  \bibinfo{journal}{Phys. Rev. B} \textbf{\bibinfo{volume}{62}},
  \bibinfo{pages}{R4790} (\bibinfo{year}{2000}).

\bibitem[{\citenamefont{Rashba}(2000)}]{RashbaPRB2000}
\bibinfo{author}{\bibfnamefont{E.~I.} \bibnamefont{Rashba}},
  \bibinfo{journal}{Phys. Rev. B} \textbf{\bibinfo{volume}{62}},
  \bibinfo{pages}{R16267} (\bibinfo{year}{2000}).

\bibitem[{\citenamefont{Smith and Silver}(2001)}]{SmithPRB2001}
\bibinfo{author}{\bibfnamefont{D.~L.} \bibnamefont{Smith}} \bibnamefont{and}
  \bibinfo{author}{\bibfnamefont{R.~N.} \bibnamefont{Silver}},
  \bibinfo{journal}{Phys. Rev. B} \textbf{\bibinfo{volume}{64}},
  \bibinfo{pages}{045323} (\bibinfo{year}{2001}).

\bibitem[{\citenamefont{Ruden and Smith}(2004)}]{RudenSmithJAP2004}
\bibinfo{author}{\bibfnamefont{P.~P.} \bibnamefont{Ruden}} \bibnamefont{and}
  \bibinfo{author}{\bibfnamefont{D.~L.} \bibnamefont{Smith}},
  \bibinfo{journal}{J. Appl. Phys.} \textbf{\bibinfo{volume}{95}},
  \bibinfo{pages}{4898} (\bibinfo{year}{2004}).

\bibitem[{\citenamefont{Albrecht and Smith}(2002)}]{SmithPRB2002}
\bibinfo{author}{\bibfnamefont{J.~D.} \bibnamefont{Albrecht}} \bibnamefont{and}
  \bibinfo{author}{\bibfnamefont{D.~L.} \bibnamefont{Smith}},
  \bibinfo{journal}{Phys. Rev. B} \textbf{\bibinfo{volume}{66}},
  \bibinfo{pages}{113303} (\bibinfo{year}{2002}).

\bibitem[{\citenamefont{Min et~al.}(2006)\citenamefont{Min, Motohashi, Lodder,
  and Jansen}}]{MinNatmat2006}
\bibinfo{author}{\bibfnamefont{B.-C.} \bibnamefont{Min}},
  \bibinfo{author}{\bibfnamefont{K.}~\bibnamefont{Motohashi}},
  \bibinfo{author}{\bibfnamefont{C.}~\bibnamefont{Lodder}}, \bibnamefont{and}
  \bibinfo{author}{\bibfnamefont{R.}~\bibnamefont{Jansen}},
  \bibinfo{journal}{Nature Material} \textbf{\bibinfo{volume}{5}},
  \bibinfo{pages}{817} (\bibinfo{year}{2006}).

\bibitem[{\citenamefont{Santos et~al.}(2007)\citenamefont{Santos, Lee, Migdal,
  Lekshmi, Satpati, and Moodera}}]{SantosPRL2007}
\bibinfo{author}{\bibfnamefont{T.~S.} \bibnamefont{Santos}},
  \bibinfo{author}{\bibfnamefont{J.~S.} \bibnamefont{Lee}},
  \bibinfo{author}{\bibfnamefont{P.}~\bibnamefont{Migdal}},
  \bibinfo{author}{\bibfnamefont{I.~C.} \bibnamefont{Lekshmi}},
  \bibinfo{author}{\bibfnamefont{B.}~\bibnamefont{Satpati}}, \bibnamefont{and}
  \bibinfo{author}{\bibfnamefont{J.~S.} \bibnamefont{Moodera}},
  \bibinfo{journal}{Phys. Rev. Lett.} \textbf{\bibinfo{volume}{98}},
  \bibinfo{pages}{016601} (\bibinfo{year}{2007}).

\bibitem[{\citenamefont{Dediu et~al.}(2008)\citenamefont{Dediu, Hueso,
  Bergenti, Riminucci, Borgatti, Graziosi, Newby, Casoli, Jong, Taliani
  et~al.}}]{DediuPRB2008}
\bibinfo{author}{\bibfnamefont{V.}~\bibnamefont{Dediu}},
  \bibinfo{author}{\bibfnamefont{L.~E.} \bibnamefont{Hueso}},
  \bibinfo{author}{\bibfnamefont{I.}~\bibnamefont{Bergenti}},
  \bibinfo{author}{\bibfnamefont{A.}~\bibnamefont{Riminucci}},
  \bibinfo{author}{\bibfnamefont{F.}~\bibnamefont{Borgatti}},
  \bibinfo{author}{\bibfnamefont{P.}~\bibnamefont{Graziosi}},
  \bibinfo{author}{\bibfnamefont{C.}~\bibnamefont{Newby}},
  \bibinfo{author}{\bibfnamefont{F.}~\bibnamefont{Casoli}},
  \bibinfo{author}{\bibfnamefont{M.~P.~D.} \bibnamefont{Jong}},
  \bibinfo{author}{\bibfnamefont{C.}~\bibnamefont{Taliani}},
  \bibnamefont{et~al.}, \bibinfo{journal}{Phys. Rev. B}
  \textbf{\bibinfo{volume}{78}}, \bibinfo{pages}{115203}
  (\bibinfo{year}{2008}).

\bibitem[{\citenamefont{Xiong et~al.}(2004)\citenamefont{Xiong, Wu, Vardeny,
  and Shi}}]{XiongNature2004}
\bibinfo{author}{\bibfnamefont{Z.~H.} \bibnamefont{Xiong}},
  \bibinfo{author}{\bibfnamefont{D.}~\bibnamefont{Wu}},
  \bibinfo{author}{\bibfnamefont{Z.~V.} \bibnamefont{Vardeny}},
  \bibnamefont{and} \bibinfo{author}{\bibfnamefont{J.}~\bibnamefont{Shi}},
  \bibinfo{journal}{Nature (London)} \textbf{\bibinfo{volume}{427}},
  \bibinfo{pages}{821} (\bibinfo{year}{2004}).

\bibitem[{\citenamefont{Wang et~al.}(2007)\citenamefont{Wang, Yang, Vardeny,
  and Li}}]{WangPRB2007}
\bibinfo{author}{\bibfnamefont{F.~J.} \bibnamefont{Wang}},
  \bibinfo{author}{\bibfnamefont{C.~G.} \bibnamefont{Yang}},
  \bibinfo{author}{\bibfnamefont{Z.~V.} \bibnamefont{Vardeny}},
  \bibnamefont{and} \bibinfo{author}{\bibfnamefont{X.~G.} \bibnamefont{Li}},
  \bibinfo{journal}{Phys. Rev. B} \textbf{\bibinfo{volume}{75}},
  \bibinfo{pages}{245324} (\bibinfo{year}{2007}).

\bibitem[{\citenamefont{Majumdara et~al.}(2006)\citenamefont{Majumdara, Laiho,
  Laukkanen, V{\"{a}}ryne, Majumdar, and {\"{O}}sterbacka}}]{OsterbackaAPL2006}
\bibinfo{author}{\bibfnamefont{S.}~\bibnamefont{Majumdara}},
  \bibinfo{author}{\bibfnamefont{R.}~\bibnamefont{Laiho}},
  \bibinfo{author}{\bibfnamefont{P.}~\bibnamefont{Laukkanen}},
  \bibinfo{author}{\bibfnamefont{I.~J.} \bibnamefont{V{\"{a}}ryne}},
  \bibinfo{author}{\bibfnamefont{H.~S.} \bibnamefont{Majumdar}},
  \bibnamefont{and}
  \bibinfo{author}{\bibfnamefont{R.}~\bibnamefont{{\"{O}}sterbacka}},
  \bibinfo{journal}{Appl. Phys. Lett.} \textbf{\bibinfo{volume}{89}},
  \bibinfo{pages}{122114} (\bibinfo{year}{2006}).

\bibitem[{\citenamefont{Wang et~al.}(2005)\citenamefont{Wang, Xiong, Wu, Shi,
  and Vardeny}}]{WangSynthMetal2005}
\bibinfo{author}{\bibfnamefont{F.~J.} \bibnamefont{Wang}},
  \bibinfo{author}{\bibfnamefont{Z.~H.} \bibnamefont{Xiong}},
  \bibinfo{author}{\bibfnamefont{D.}~\bibnamefont{Wu}},
  \bibinfo{author}{\bibfnamefont{J.}~\bibnamefont{Shi}}, \bibnamefont{and}
  \bibinfo{author}{\bibfnamefont{Z.~V.} \bibnamefont{Vardeny}},
  \bibinfo{journal}{Synthetic Metals} \textbf{\bibinfo{volume}{155}},
  \bibinfo{pages}{172} (\bibinfo{year}{2005}).

\bibitem[{\citenamefont{Pramanik et~al.}(2006)\citenamefont{Pramanik,
  Bandyopadhyay, Garre, and Cahay}}]{PramanikPRB2006}
\bibinfo{author}{\bibfnamefont{S.}~\bibnamefont{Pramanik}},
  \bibinfo{author}{\bibfnamefont{S.}~\bibnamefont{Bandyopadhyay}},
  \bibinfo{author}{\bibfnamefont{K.}~\bibnamefont{Garre}}, \bibnamefont{and}
  \bibinfo{author}{\bibfnamefont{M.}~\bibnamefont{Cahay}},
  \bibinfo{journal}{Phys. Rev. B} \textbf{\bibinfo{volume}{74}},
  \bibinfo{pages}{235329} (\bibinfo{year}{2006}).

\bibitem[{\citenamefont{Xu et~al.}(2007)\citenamefont{Xu, Szulczewski, LeClair,
  Navarrete, Schad, Miao, Guo, and Gupta}}]{XuAPL2007}
\bibinfo{author}{\bibfnamefont{W.}~\bibnamefont{Xu}},
  \bibinfo{author}{\bibfnamefont{G.~J.} \bibnamefont{Szulczewski}},
  \bibinfo{author}{\bibfnamefont{P.}~\bibnamefont{LeClair}},
  \bibinfo{author}{\bibfnamefont{I.}~\bibnamefont{Navarrete}},
  \bibinfo{author}{\bibfnamefont{R.}~\bibnamefont{Schad}},
  \bibinfo{author}{\bibfnamefont{G.}~\bibnamefont{Miao}},
  \bibinfo{author}{\bibfnamefont{H.}~\bibnamefont{Guo}}, \bibnamefont{and}
  \bibinfo{author}{\bibfnamefont{A.}~\bibnamefont{Gupta}},
  \bibinfo{journal}{Appl. Phys. Lett.} \textbf{\bibinfo{volume}{90}},
  \bibinfo{pages}{072506} (\bibinfo{year}{2007}).

\bibitem[{\citenamefont{Petta et~al.}(2004)\citenamefont{Petta, Slater, and
  Ralph}}]{PettaPRL2004}
\bibinfo{author}{\bibfnamefont{J.~R.} \bibnamefont{Petta}},
  \bibinfo{author}{\bibfnamefont{S.~K.} \bibnamefont{Slater}},
  \bibnamefont{and} \bibinfo{author}{\bibfnamefont{D.~C.} \bibnamefont{Ralph}},
  \bibinfo{journal}{Phys. Rev. Lett.} \textbf{\bibinfo{volume}{93}},
  \bibinfo{pages}{136601} (\bibinfo{year}{2004}).

\bibitem[{\citenamefont{Ikegami et~al.}(2008)\citenamefont{Ikegami, Kawayama,
  Tonouchi, Nakao, Yamashita, and Tada1}}]{IkegamiAPL2008}
\bibinfo{author}{\bibfnamefont{T.}~\bibnamefont{Ikegami}},
  \bibinfo{author}{\bibfnamefont{I.}~\bibnamefont{Kawayama}},
  \bibinfo{author}{\bibfnamefont{M.}~\bibnamefont{Tonouchi}},
  \bibinfo{author}{\bibfnamefont{S.}~\bibnamefont{Nakao}},
  \bibinfo{author}{\bibfnamefont{Y.}~\bibnamefont{Yamashita}},
  \bibnamefont{and} \bibinfo{author}{\bibfnamefont{H.}~\bibnamefont{Tada1}},
  \bibinfo{journal}{Appl. Phys. Lett.} \textbf{\bibinfo{volume}{92}},
  \bibinfo{pages}{153304} (\bibinfo{year}{2008}).

\bibitem[{\citenamefont{Jiang et~al.}(2008)\citenamefont{Jiang, Pearson, and
  Bader}}]{JiangPRB2008}
\bibinfo{author}{\bibfnamefont{J.~S.} \bibnamefont{Jiang}},
  \bibinfo{author}{\bibfnamefont{J.~E.} \bibnamefont{Pearson}},
  \bibnamefont{and} \bibinfo{author}{\bibfnamefont{S.~D.} \bibnamefont{Bader}},
  \bibinfo{journal}{Phys. Rev. B} \textbf{\bibinfo{volume}{77}},
  \bibinfo{pages}{035303} (\bibinfo{year}{2008}).

\bibitem[{\citenamefont{Vinzelberg et~al.}(2008)\citenamefont{Vinzelberg,
  Schumann, Elefant, Gangineni, Thomas, and B{\"{u}}chner}}]{Vinzelberg2008}
\bibinfo{author}{\bibfnamefont{H.}~\bibnamefont{Vinzelberg}},
  \bibinfo{author}{\bibfnamefont{J.}~\bibnamefont{Schumann}},
  \bibinfo{author}{\bibfnamefont{D.}~\bibnamefont{Elefant}},
  \bibinfo{author}{\bibfnamefont{R.~B.} \bibnamefont{Gangineni}},
  \bibinfo{author}{\bibfnamefont{J.}~\bibnamefont{Thomas}}, \bibnamefont{and}
  \bibinfo{author}{\bibfnamefont{B.}~\bibnamefont{B{\"{u}}chner}},
  \bibinfo{journal}{J. Appl. Phys.} \textbf{\bibinfo{volume}{103}},
  \bibinfo{pages}{093720} (\bibinfo{year}{2008}).

\bibitem[{\citenamefont{Fullerton et~al.}(1992)\citenamefont{Fullerton, Kelly,
  Guimpel, Schuller, and Bruynseraede}}]{FullertonPRL1992}
\bibinfo{author}{\bibfnamefont{E.~E.} \bibnamefont{Fullerton}},
  \bibinfo{author}{\bibfnamefont{D.~M.} \bibnamefont{Kelly}},
  \bibinfo{author}{\bibfnamefont{J.}~\bibnamefont{Guimpel}},
  \bibinfo{author}{\bibfnamefont{I.~K.} \bibnamefont{Schuller}},
  \bibnamefont{and}
  \bibinfo{author}{\bibfnamefont{Y.}~\bibnamefont{Bruynseraede}},
  \bibinfo{journal}{Phys. Rev. Lett.} \textbf{\bibinfo{volume}{68}},
  \bibinfo{pages}{859} (\bibinfo{year}{1992}).

\bibitem[{\citenamefont{Schad et~al.}(1998)\citenamefont{Schad, Beli{\"{e}}n,
  Verbanck, Potter, Fischer, Lefebvre, Bessiere, Moshchalkov, and
  Bruynseraede}}]{SchadPRB1998}
\bibinfo{author}{\bibfnamefont{R.}~\bibnamefont{Schad}},
  \bibinfo{author}{\bibfnamefont{P.}~\bibnamefont{Beli{\"{e}}n}},
  \bibinfo{author}{\bibfnamefont{G.}~\bibnamefont{Verbanck}},
  \bibinfo{author}{\bibfnamefont{C.~D.} \bibnamefont{Potter}},
  \bibinfo{author}{\bibfnamefont{H.}~\bibnamefont{Fischer}},
  \bibinfo{author}{\bibfnamefont{S.}~\bibnamefont{Lefebvre}},
  \bibinfo{author}{\bibfnamefont{M.}~\bibnamefont{Bessiere}},
  \bibinfo{author}{\bibfnamefont{V.~V.} \bibnamefont{Moshchalkov}},
  \bibnamefont{and}
  \bibinfo{author}{\bibfnamefont{Y.}~\bibnamefont{Bruynseraede}},
  \bibinfo{journal}{Phys. Rev. B} \textbf{\bibinfo{volume}{57}},
  \bibinfo{pages}{13692} (\bibinfo{year}{1998}).

\bibitem[{\citenamefont{Fitzsimmons et~al.}(2004)\citenamefont{Fitzsimmons,
  Bader, Borchers, Felcher, Furdyna, Hoffman, Kortright, Schuller, Schulthess,
  Sinha et~al.}}]{Fitz2004}
\bibinfo{author}{\bibfnamefont{M.~R.} \bibnamefont{Fitzsimmons}},
  \bibinfo{author}{\bibfnamefont{S.~D.} \bibnamefont{Bader}},
  \bibinfo{author}{\bibfnamefont{J.~A.} \bibnamefont{Borchers}},
  \bibinfo{author}{\bibfnamefont{G.~P.} \bibnamefont{Felcher}},
  \bibinfo{author}{\bibfnamefont{J.~K.} \bibnamefont{Furdyna}},
  \bibinfo{author}{\bibfnamefont{A.}~\bibnamefont{Hoffman}},
  \bibinfo{author}{\bibfnamefont{J.~B.} \bibnamefont{Kortright}},
  \bibinfo{author}{\bibfnamefont{I.~K.} \bibnamefont{Schuller}},
  \bibinfo{author}{\bibfnamefont{T.~C.} \bibnamefont{Schulthess}},
  \bibinfo{author}{\bibfnamefont{S.~K.} \bibnamefont{Sinha}},
  \bibnamefont{et~al.}, \bibinfo{journal}{J. Mag. Magn. Mater.}
  \textbf{\bibinfo{volume}{271}}, \bibinfo{pages}{103} (\bibinfo{year}{2004}).

\bibitem[{\citenamefont{Majkrzak}(1991)}]{MajkrzakPhysicaB1991}
\bibinfo{author}{\bibfnamefont{C.~F.} \bibnamefont{Majkrzak}},
  \bibinfo{journal}{Physica B} \textbf{\bibinfo{volume}{173}},
  \bibinfo{pages}{75} (\bibinfo{year}{1991}).

\bibitem[{\citenamefont{Fitzsimmons and Majkrzak}(2005)}]{Fitz2005}
\bibinfo{author}{\bibfnamefont{M.~R.} \bibnamefont{Fitzsimmons}}
  \bibnamefont{and} \bibinfo{author}{\bibfnamefont{C.}~\bibnamefont{Majkrzak}},
  \emph{\bibinfo{title}{Modern Techniques for Characterizing Magnetic
  Materials}} (\bibinfo{publisher}{Springer}, \bibinfo{year}{2005}),
  chap.~\bibinfo{chapter}{3}, pp. \bibinfo{pages}{107--155}.

\bibitem[{\citenamefont{Majkrzak et~al.}(1988)\citenamefont{Majkrzak, Gibbs,
  B\"{o}ni, Goldman, Kwo, Hong, Hsieh, Fleming, McWhan, Yafet
  et~al.}}]{MajkrzakJAP1988}
\bibinfo{author}{\bibfnamefont{C.~F.} \bibnamefont{Majkrzak}},
  \bibinfo{author}{\bibfnamefont{D.}~\bibnamefont{Gibbs}},
  \bibinfo{author}{\bibfnamefont{P.}~\bibnamefont{B\"{o}ni}},
  \bibinfo{author}{\bibfnamefont{A.~I.} \bibnamefont{Goldman}},
  \bibinfo{author}{\bibfnamefont{J.}~\bibnamefont{Kwo}},
  \bibinfo{author}{\bibfnamefont{M.}~\bibnamefont{Hong}},
  \bibinfo{author}{\bibfnamefont{T.~C.} \bibnamefont{Hsieh}},
  \bibinfo{author}{\bibfnamefont{R.~M.} \bibnamefont{Fleming}},
  \bibinfo{author}{\bibfnamefont{D.~B.} \bibnamefont{McWhan}},
  \bibinfo{author}{\bibfnamefont{Y.}~\bibnamefont{Yafet}},
  \bibnamefont{et~al.}, \bibinfo{journal}{J. Appl. Phys.}
  \textbf{\bibinfo{volume}{63}}, \bibinfo{pages}{3447} (\bibinfo{year}{1988}).

\bibitem[{\citenamefont{Kienzle et~al.}()\citenamefont{Kienzle, O'Donovan,
  Ankner, Berk, and Majkrzak}}]{reflpak}
\bibinfo{author}{\bibfnamefont{P.~A.} \bibnamefont{Kienzle}},
  \bibinfo{author}{\bibfnamefont{K.~V.} \bibnamefont{O'Donovan}},
  \bibinfo{author}{\bibfnamefont{J.~F.} \bibnamefont{Ankner}},
  \bibinfo{author}{\bibfnamefont{N.~F.} \bibnamefont{Berk}}, \bibnamefont{and}
  \bibinfo{author}{\bibfnamefont{C.~F.} \bibnamefont{Majkrzak}},
  \bibinfo{note}{\href{http://www.ncnr.nist.gov/reflpak}{http://www.ncnr.nist.%
gov/reflpak}, 2000-2006}.

\bibitem[{\citenamefont{Zhan et~al.}(2008)\citenamefont{Zhan, de~Jong, Li,
  Dediu, Fahlman, and Salaneck}}]{ZhanPRB2008}
\bibinfo{author}{\bibfnamefont{Y.~Q.} \bibnamefont{Zhan}},
  \bibinfo{author}{\bibfnamefont{M.~P.} \bibnamefont{de~Jong}},
  \bibinfo{author}{\bibfnamefont{F.~H.} \bibnamefont{Li}},
  \bibinfo{author}{\bibfnamefont{V.}~\bibnamefont{Dediu}},
  \bibinfo{author}{\bibfnamefont{M.}~\bibnamefont{Fahlman}}, \bibnamefont{and}
  \bibinfo{author}{\bibfnamefont{W.~R.} \bibnamefont{Salaneck}},
  \bibinfo{journal}{Phys. Rev. B} \textbf{\bibinfo{volume}{78}},
  \bibinfo{eid}{045208} (pages~\bibinfo{numpages}{6}) (\bibinfo{year}{2008}).

\bibitem[{\citenamefont{J\"{o}nsson-{\AA}kerman
  et~al.}(2000)\citenamefont{J\"{o}nsson-{\AA}kerman, Escudero, Leighton, Kim,
  Schuller, and Rabson}}]{AkermanAPL2000}
\bibinfo{author}{\bibfnamefont{B.~J.} \bibnamefont{J\"{o}nsson-{\AA}kerman}},
  \bibinfo{author}{\bibfnamefont{R.}~\bibnamefont{Escudero}},
  \bibinfo{author}{\bibfnamefont{C.}~\bibnamefont{Leighton}},
  \bibinfo{author}{\bibfnamefont{S.}~\bibnamefont{Kim}},
  \bibinfo{author}{\bibfnamefont{I.~K.} \bibnamefont{Schuller}},
  \bibnamefont{and} \bibinfo{author}{\bibfnamefont{D.~A.}
  \bibnamefont{Rabson}}, \bibinfo{journal}{Appl. Phys. Lett.}
  \textbf{\bibinfo{volume}{77}}, \bibinfo{pages}{1870} (\bibinfo{year}{2000}).

\bibitem[{\citenamefont{Julliere}(1975)}]{Julliere1975}
\bibinfo{author}{\bibfnamefont{M.}~\bibnamefont{Julliere}},
  \bibinfo{journal}{Phys. Lett. A} \textbf{\bibinfo{volume}{54}},
  \bibinfo{pages}{225} (\bibinfo{year}{1975}).

\bibitem[{\citenamefont{Pramanik et~al.}(2007)\citenamefont{Pramanik,
  Stefanita, Patibandla, Bandyopadhyay, Garre, Harth, and
  Cahay}}]{PramanikNatNano2007}
\bibinfo{author}{\bibfnamefont{S.}~\bibnamefont{Pramanik}},
  \bibinfo{author}{\bibfnamefont{C.~G.} \bibnamefont{Stefanita}},
  \bibinfo{author}{\bibfnamefont{S.}~\bibnamefont{Patibandla}},
  \bibinfo{author}{\bibfnamefont{S.}~\bibnamefont{Bandyopadhyay}},
  \bibinfo{author}{\bibfnamefont{K.}~\bibnamefont{Garre}},
  \bibinfo{author}{\bibfnamefont{N.}~\bibnamefont{Harth}}, \bibnamefont{and}
  \bibinfo{author}{\bibfnamefont{M.}~\bibnamefont{Cahay}},
  \bibinfo{journal}{Nature Nanotechnology} \textbf{\bibinfo{volume}{2}},
  \bibinfo{pages}{216} (\bibinfo{year}{2007}).

\bibitem[{\citenamefont{Shim et~al.}(2008)\citenamefont{Shim, Raman, Park,
  Santos, Miao, Satpati, and Moodera}}]{ShimPRL2008}
\bibinfo{author}{\bibfnamefont{J.~H.} \bibnamefont{Shim}},
  \bibinfo{author}{\bibfnamefont{K.~V.} \bibnamefont{Raman}},
  \bibinfo{author}{\bibfnamefont{Y.~J.} \bibnamefont{Park}},
  \bibinfo{author}{\bibfnamefont{T.~S.} \bibnamefont{Santos}},
  \bibinfo{author}{\bibfnamefont{G.~X.} \bibnamefont{Miao}},
  \bibinfo{author}{\bibfnamefont{B.}~\bibnamefont{Satpati}}, \bibnamefont{and}
  \bibinfo{author}{\bibfnamefont{J.~S.} \bibnamefont{Moodera}},
  \bibinfo{journal}{Phys. Rev. Lett.}  (\bibinfo{year}{2008}).

\bibitem[{\citenamefont{Strijkers et~al.}(2001)\citenamefont{Strijkers, Ji,
  Yang, Chien, and Byers}}]{StrijkersPRB2001}
\bibinfo{author}{\bibfnamefont{G.~J.} \bibnamefont{Strijkers}},
  \bibinfo{author}{\bibfnamefont{Y.}~\bibnamefont{Ji}},
  \bibinfo{author}{\bibfnamefont{F.~Y.} \bibnamefont{Yang}},
  \bibinfo{author}{\bibfnamefont{C.~L.} \bibnamefont{Chien}}, \bibnamefont{and}
  \bibinfo{author}{\bibfnamefont{J.~M.} \bibnamefont{Byers}},
  \bibinfo{journal}{Phys. Rev. B} \textbf{\bibinfo{volume}{63}},
  \bibinfo{pages}{104510} (\bibinfo{year}{2001}).

\bibitem[{\citenamefont{Ankner and Majkrzak}(1992)}]{MajkrzakSPIE1992}
\bibinfo{author}{\bibfnamefont{J.~F.} \bibnamefont{Ankner}} \bibnamefont{and}
  \bibinfo{author}{\bibfnamefont{C.~F.} \bibnamefont{Majkrzak}}, in
  \emph{\bibinfo{booktitle}{Neutron Optical Devices and Applications}}, edited
  by \bibinfo{editor}{\bibfnamefont{C.~F.} \bibnamefont{Majkrzak}}
  \bibnamefont{and} \bibinfo{editor}{\bibfnamefont{J.~L.} \bibnamefont{Wood}}
  (\bibinfo{publisher}{SPIE}, \bibinfo{year}{1992}), vol.
  \bibinfo{volume}{1738}, pp. \bibinfo{pages}{260--269}.

\bibitem[{\citenamefont{Smith}(1994)}]{ElecSpec1994}
\bibinfo{author}{\bibfnamefont{G.~C.} \bibnamefont{Smith}},
  \emph{\bibinfo{title}{Surface analysis by electron spectroscopy ¡ª
  Measurement and interpretation}} (\bibinfo{publisher}{Plenum Press},
  \bibinfo{year}{1994}), chap.~\bibinfo{chapter}{6}, pp.
  \bibinfo{pages}{75--85}.

\bibitem[{\citenamefont{Cable et~al.}(1986)\citenamefont{Cable, Khan, Felcher,
  and Schuller}}]{CablePRB1986}
\bibinfo{author}{\bibfnamefont{J.~W.} \bibnamefont{Cable}},
  \bibinfo{author}{\bibfnamefont{M.~R.} \bibnamefont{Khan}},
  \bibinfo{author}{\bibfnamefont{G.~P.} \bibnamefont{Felcher}},
  \bibnamefont{and} \bibinfo{author}{\bibfnamefont{I.~K.}
  \bibnamefont{Schuller}}, \bibinfo{journal}{Phys. Rev. B}
  \textbf{\bibinfo{volume}{34}}, \bibinfo{pages}{1643} (\bibinfo{year}{1986}).

\bibitem[{\citenamefont{Pechan et~al.}(1994)\citenamefont{Pechan, Ankner,
  Majkrazk, Kelly, and Schuller}}]{PechanJAP1994}
\bibinfo{author}{\bibfnamefont{M.~J.} \bibnamefont{Pechan}},
  \bibinfo{author}{\bibfnamefont{J.~F.} \bibnamefont{Ankner}},
  \bibinfo{author}{\bibfnamefont{C.~F.} \bibnamefont{Majkrazk}},
  \bibinfo{author}{\bibfnamefont{D.~M.} \bibnamefont{Kelly}}, \bibnamefont{and}
  \bibinfo{author}{\bibfnamefont{I.~K.} \bibnamefont{Schuller}},
  \bibinfo{journal}{J. Appl. Phys.} \textbf{\bibinfo{volume}{75}},
  \bibinfo{pages}{6178} (\bibinfo{year}{1994}).

\bibitem[{\citenamefont{Hoffmann et~al.}(2005)\citenamefont{Hoffmann,
  te~Velthuis, Sefrioui, Santamar\'{\i}a, Fitzsimmons, Park, and
  Varela}}]{HoffmannPRB2005}
\bibinfo{author}{\bibfnamefont{A.}~\bibnamefont{Hoffmann}},
  \bibinfo{author}{\bibfnamefont{S.~G.~E.} \bibnamefont{te~Velthuis}},
  \bibinfo{author}{\bibfnamefont{Z.}~\bibnamefont{Sefrioui}},
  \bibinfo{author}{\bibfnamefont{J.}~\bibnamefont{Santamar\'{\i}a}},
  \bibinfo{author}{\bibfnamefont{M.~R.} \bibnamefont{Fitzsimmons}},
  \bibinfo{author}{\bibfnamefont{S.}~\bibnamefont{Park}}, \bibnamefont{and}
  \bibinfo{author}{\bibfnamefont{M.}~\bibnamefont{Varela}},
  \bibinfo{journal}{Phys. Rev. B} \textbf{\bibinfo{volume}{72}},
  \bibinfo{eid}{140407} (\bibinfo{year}{2005}).

\bibitem[{\citenamefont{Borchers et~al.}(1996)\citenamefont{Borchers, Gehring,
  Erwin, Ankner, Majkrzak, Hylton, Coffey, Parker, and
  Howard}}]{BorchersPRB1996}
\bibinfo{author}{\bibfnamefont{J.~A.} \bibnamefont{Borchers}},
  \bibinfo{author}{\bibfnamefont{P.~M.} \bibnamefont{Gehring}},
  \bibinfo{author}{\bibfnamefont{R.~W.} \bibnamefont{Erwin}},
  \bibinfo{author}{\bibfnamefont{J.~F.} \bibnamefont{Ankner}},
  \bibinfo{author}{\bibfnamefont{C.~F.} \bibnamefont{Majkrzak}},
  \bibinfo{author}{\bibfnamefont{T.~L.} \bibnamefont{Hylton}},
  \bibinfo{author}{\bibfnamefont{K.~R.} \bibnamefont{Coffey}},
  \bibinfo{author}{\bibfnamefont{M.~A.} \bibnamefont{Parker}},
  \bibnamefont{and} \bibinfo{author}{\bibfnamefont{J.~K.}
  \bibnamefont{Howard}}, \bibinfo{journal}{Phys. Rev. B}
  \textbf{\bibinfo{volume}{54}}, \bibinfo{pages}{9870} (\bibinfo{year}{1996}).

\bibitem[{\citenamefont{Borchers et~al.}(1999)\citenamefont{Borchers, Dura,
  Unguris, Tulchinsky, Kelley, Majkrzak, Hsu, Loloee, Pratt, and
  Bass}}]{BorchersPRL1999}
\bibinfo{author}{\bibfnamefont{J.~A.} \bibnamefont{Borchers}},
  \bibinfo{author}{\bibfnamefont{J.~A.} \bibnamefont{Dura}},
  \bibinfo{author}{\bibfnamefont{J.}~\bibnamefont{Unguris}},
  \bibinfo{author}{\bibfnamefont{D.}~\bibnamefont{Tulchinsky}},
  \bibinfo{author}{\bibfnamefont{M.~H.} \bibnamefont{Kelley}},
  \bibinfo{author}{\bibfnamefont{C.~F.} \bibnamefont{Majkrzak}},
  \bibinfo{author}{\bibfnamefont{S.~Y.} \bibnamefont{Hsu}},
  \bibinfo{author}{\bibfnamefont{R.}~\bibnamefont{Loloee}},
  \bibinfo{author}{\bibfnamefont{W.~P.} \bibnamefont{Pratt}}, \bibnamefont{and}
  \bibinfo{author}{\bibfnamefont{J.}~\bibnamefont{Bass}},
  \bibinfo{journal}{Phys. Rev. Lett.} \textbf{\bibinfo{volume}{82}},
  \bibinfo{pages}{2796} (\bibinfo{year}{1999}).

\bibitem[{\citenamefont{Scott}(2003)}]{ScottVacSci2003}
\bibinfo{author}{\bibfnamefont{J.~C.} \bibnamefont{Scott}},
  \bibinfo{journal}{J. Vac. Sci. Technol A} \textbf{\bibinfo{volume}{21}},
  \bibinfo{pages}{521} (\bibinfo{year}{2003}).

\bibitem[{\citenamefont{Brinkman et~al.}(1970)\citenamefont{Brinkman, Dynes,
  and Rowell}}]{BDRJAP1970}
\bibinfo{author}{\bibfnamefont{W.~F.} \bibnamefont{Brinkman}},
  \bibinfo{author}{\bibfnamefont{R.~C.} \bibnamefont{Dynes}}, \bibnamefont{and}
  \bibinfo{author}{\bibfnamefont{J.~M.} \bibnamefont{Rowell}},
  \bibinfo{journal}{J. Appl. Phys.} \textbf{\bibinfo{volume}{41}},
  \bibinfo{pages}{1915} (\bibinfo{year}{1970}).

\bibitem[{\citenamefont{Simmons}(1963)}]{SimmonsJAP1963}
\bibinfo{author}{\bibfnamefont{J.~G.} \bibnamefont{Simmons}},
  \bibinfo{journal}{J. Appl. Phys.} \textbf{\bibinfo{volume}{34}},
  \bibinfo{pages}{2581} (\bibinfo{year}{1963}).

\bibitem[{\citenamefont{Simmons}(1964)}]{SimmonsJAP1964}
\bibinfo{author}{\bibfnamefont{J.~G.} \bibnamefont{Simmons}},
  \bibinfo{journal}{J. Appl. Phys.} \textbf{\bibinfo{volume}{35}},
  \bibinfo{pages}{2655} (\bibinfo{year}{1964}).

\bibitem[{\citenamefont{Campbell and Crone}(2007)}]{CambellAPL2007}
\bibinfo{author}{\bibfnamefont{I.~H.} \bibnamefont{Campbell}} \bibnamefont{and}
  \bibinfo{author}{\bibfnamefont{B.~K.} \bibnamefont{Crone}},
  \bibinfo{journal}{Appl. Phys. Lett.} \textbf{\bibinfo{volume}{90}},
  \bibinfo{pages}{242107} (\bibinfo{year}{2007}).

\bibitem[{\citenamefont{Rudner and Levitov}(2007)}]{RudnerPRL2007}
\bibinfo{author}{\bibfnamefont{M.~S.} \bibnamefont{Rudner}} \bibnamefont{and}
  \bibinfo{author}{\bibfnamefont{L.~S.} \bibnamefont{Levitov}},
  \bibinfo{journal}{Phys. Rev. Lett.} \textbf{\bibinfo{volume}{99}},
  \bibinfo{eid}{246602} (\bibinfo{year}{2007}).

\end{thebibliography}

\end{document}